\documentclass[twocolumn, pra,showpacs,nofootinbib]{revtex4}

\usepackage{dcolumn,graphicx}
\begin{document}
\preprint{UNR September 2002-\today }
\title{ Anisotropic pseudo-potential for polarized dilute quantum gases}
\author{Andrei Derevianko}
\email{andrei@physics.unr.edu}
\affiliation { Department of Physics, University of Nevada, Reno, Nevada 89557}
\date{\today}

\begin{abstract}
Anisotropic pseudopotential
relevant to collisions of two particles polarized by
external field is rigorously derived and its properties are investigated.
Such low-energy pseudopotential may be useful in describing
collective properties of dilute quantum gases, such as molecules polarized
by electric field or metastable $^3P_2$ atoms polarized by
magnetic field.
The pseudopotential is  expressed in terms of reactance (K--) matrix and
derivatives of Dirac delta-function.
In most applications it may be represented
as a sum of traditional spherically-symmetric contact term and
anisotropic part. The former contribution may
be parameterized by a generalized scattering length.
The anisotropic part of pseudopotential may be characterized by
off-diagonal scattering length for dipolar
interactions and off-diagonal scattering volume for quadrupolar
interactions. Two-body matrix element
of the pseudopotential in a basis of plane waves is also derived.
\end{abstract}

\maketitle

\section{Introduction and problem setup}
\label{Sec_intro}
The concept of pseudo-potential, i.e. full interparticle interaction being replaced by
some less complicated ``effective'' potential, plays an important role in many
subfields of physics~\cite{DemOst88}. In particular, properties of traditional
Bose-Einstein condensates (BEC)
may be well understood just in terms of a delta-function potential, with
its strength determined by a single parameter -- scattering length \cite{DalGioPit99,Leg01}
which characterizes low-energy scattering between two particles.
A rigorous derivation of pseudo-potential for spherically-symmetric
interactions has been carried out by \citet{HuaYan57}.
Here I extend their derivation to
{\em anisotropic} interactions.
I also evaluate a matrix element of the derived anisotropic pseudo-potential in the basis
of plane waves;
this matrix element may be useful in studies of many-body properties
of quantum degenerate gasses.

Unusual collective properties of bosons and fermions with
anisotropic interactions have generated a considerable interest over the last few
years, see e.g. Refs.~\cite{YiYou00,YiYou02,GioGorPfa02,SanShlZol00,GorRzaPfa00,%
GorSan02,GorSanLew02,GioODeKur02,KrzEngRze01,BarMarRyc02}.
Below I enumerate several
systems where the results of my analysis may be applicable.
First, \citet{YiYou00} considered an application of strong electric field to an
atomic condensate. The electric field induces atomic electric dipoles and thus anisotropic
dipole-dipole interactions between the atoms. Another novel systems where the anisotropic
interactions
dominate at large separations are  heteronuclear molecules\cite{SanShlZol00}.
Here an application of electric field is required to freeze the rotations of
the molecules and to align the intrinsic molecular dipole moments with the
field. {\em Magnetic} dipole-dipole interactions are present even
for well-studied alkali-metal atoms. These interactions
may be amplified for more complex atoms like europium and
chromium~\cite{KimFriKat97,WeideCKim98,BelStuLoc99} with larger
magnetic momenta of the ground atomic state. The influence of such magnetic
dipolar interactions on the condensate properties was discussed in
Ref.~\cite{GorRzaPfa00}. New systems where the anisotropy of interactions
may be also of interest are metastable $^{3}\!P_{2}$ alkaline-earth
atoms placed in external magnetic field. Here the long-range forces are
dominated by interactions of atomic quadrupoles~\cite{DerPorKot03}.
It should be noted that the application of external magnetic or electric
field is important in all these
examples - the field fixes quantization axis and a condensate may be described in terms
of a single order parameter.

\begin{figure}[h]
\begin{center}
\includegraphics*[scale=0.75]{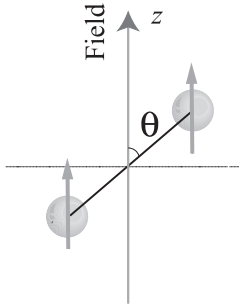}
\caption{ Geometry of collision process.
At large separations colliding particles are polarized  along external field.
During the collision the particles are assumed to follow a unique adiabatic potential.
 \label{Fig_polAtoms}}
\end{center}
\end{figure}

In all the enumerated examples the collision process may be formalized using
Fig.~\ref{Fig_polAtoms}. Here we show a pair of identical particles interacting in the presence of
external uniform field.
The $z$-axis is chosen along the direction of the field and angle $\theta$
determines orientation of collision (interatomic) axis $\hat\mathbf{r}$
with respect to the field.
At large separations $r$ the particles are polarized along the direction of the
field. In the most general case, as a result of a collision, a
change in polarization may occur (e.g. dipole moment of a molecule could end up pointing
in the direction opposite to the field). We will disregard these non-adiabatic
collisions. Then the interaction between the particles may be described by
a unique  potential $V(r,\theta)$. Without loss of generality,
this axially-symmetric potential may be expanded into Legendre polynomials
$P_L(\cos \theta)$
\begin{eqnarray}
V\left(r, \theta \right) &=&
V_\mathrm{sph}\left( r \right)  + V_\mathrm{anis}\left( r, \theta \right)\, ,  \\
V_\mathrm{anis}\left( r, \theta \right) &=& \sum_{L=2,4,..} V_L(r) \,
P_{L}\left(  \cos\theta\right)  \, .\label{Eq_Vexpand}%
\end{eqnarray}
Here $V_\mathrm{sph}\left( r \right)$ and $V_\mathrm{anis}\left( r, \theta \right)$
are spherically-symmetric ($L=0$) and non-spherical contributions respectively.
Although all even $L$ contribute,
at sufficiently large $r$ the anisotropic contribution may be dominated
by a single $L$ term. In particular, we will focus on two practically interesting cases -
dipolar ($L=2$)
\begin{equation}
V_\mathrm{anis}\left( r, \theta \right)  \rightarrow V_\mathrm{DD} =
\frac{C_3}{r^{3}}%
P_{2}\left(  \cos\theta\right)  \, , r \rightarrow \infty  \,,\label{Eq_VDD}%
\end{equation}
and quadrupolar ($L=4$)
\begin{equation}
V_\mathrm{anis}\left( r, \theta \right)
  \rightarrow V_\mathrm{QQ} = \frac{C_5}{r^{5}}%
P_{4}\left(  \cos\theta\right) \, , r \rightarrow \infty  \label{Eq_VQQ}%
\end{equation}
interactions. In the above equations constants $C_{L+1}$ parameterize
strengths of interactions and are proportional to the squares of respective multipole
moments (e.g. molecular dipole moment).

Realistic
interaction potentials $V(\mathbf{r})$ are singular at small
interparticle separations and this singularity leads to well-known problems~\cite{FetWal71}
in formulating perturbative expansion for many-body properties.
In particular,
matrix element of the interaction in basis of free-particle is divergent.
To remedy this problem,
the full interaction potential $V$ is usually replaced by a pseudopotential
$\hat{\mathcal V}$.
A rigorous derivation of pseudo-potential for spherically-symmetric
interactions has been carried out by \citet{HuaYan57}. Here I extend their method
to anisotropic interactions.

Previously, for anisotropic
dipolar interactions, \citet{YiYou00} proposed the following pseudopotential
\begin{equation}
\hat{\mathcal V}^{\rm YY}_{\rm DD} = g \delta(\mathbf {r} ) +
 \frac{C_3}{r^{3}} \,
P_{2}\left(  \cos\theta\right)  \, .
\label{Eqn_YY}
\end{equation}
Here the first term is related to spherically-symmetric part of
the full potential~(\ref{Eq_Vexpand}) and the second contribution is simply
the long-range dipolar interaction~(\ref{Eq_VDD}).
This pseudopotential has been employed in a large number of studies
of properties of BECs with dipolar
interactions, see, e.g.~\cite{YiYou00,YiYou02,GioGorPfa02,SanShlZol00,GorRzaPfa00,GorSan02,%
GorSanLew02,GioODeKur02}.
Although straightforward to work with in applications,
the pseudo-potential (\ref{Eqn_YY}) has certain shortcomings.
For example, it is not valid in the vicinity of resonances.
Pseudo-potential derived here resolves these shortcomings.


The goal of this work is to consistently develop a pseudo-potential method for
non-spherical interaction potentials. It will be required that
two-body wavefunctions obtained
with the pseudo-potential $\hat{\mathcal V}$ and full original potential $V$ to be equal at large interparticle
separations. The derivation of this pseudo-potential is carried out
in Section~\ref{Sec_APP}. Certain properties of the derived pseudopotential
are discussed in Section~\ref{Sec_prop} and
we specialize the discussion to dipolar and quadrupolar
interactions in Section~\ref{Sec_DQ}. Matrix element of the pseudo-potential in free-particle
basis is evaluated in Section~\ref{Sec_mel}. Finally, Appendix contains a derivation
of certain low-energy limits of K-matrix in the Born approximation.

\section{ Anisotropic pseudo-potential }
\label{Sec_APP}
We consider a solution of the Schr\"{o}dinger equation for a relative motion of
two particles interacting through a potential $V(\mathbf{r})$%
\begin{equation}
\frac{\hbar^{2}}{2\mu}\left(  \mathbf{\nabla}^{2}+k^{2}\right)  \Phi\left(
\mathbf{r}\right)  =V\left(  \mathbf{r}\right)  \Phi\left(  \mathbf{r}\right) \, ,
\label{Eq_SE}
\end{equation}
where $\mu$ is reduced mass of the pair and $\mathbf{k}$ is the relative
linear momentum. We assume that at sufficiently large separations $r>r_{c},$
$r^{2}V\left(  \mathbf{r}\right)  \rightarrow0$ for any direction~$\hat{r}$.
We also assume that the particles are contained in some large volume with
characteristic size much larger than the extent of the potential $r_{c}$. Some
arbitrary boundary conditions may be imposed on the surface of the enclosing
volume. At $r\gg r_{c}$ the wavefunction $\Phi\left(  \mathbf{r}\right)  $ may
be expanded in free-particle solutions
\begin{equation}
\Phi_{\infty}\left(  \mathbf{r}%
\right)  =\sum_{lm}\left(  \alpha_{lm}\, j_{l}\left(  kr\right)  -\beta
_{lm}n_{l}\left(  kr\right)  \right)  Y_{lm}\left(  \theta,\varphi\right)
\, ,\label{Eqn_asmpt}%
\end{equation}
where $j_{l}\left(  kr\right)  $ and $n_{l}\left(  kr\right)  $ are spherical
Bessel and Neumann functions respectively and $\alpha_{lm}$ and $\beta_{lm}$
are integration constants.

Following Ref.~\cite{HuaYan57} the pseudo-potential is determined by acting
with $\frac{\hbar^{2}}{2\mu}\left(  \mathbf{\nabla}^{2}+k^{2}\right)  $ on the
asymptotic form (\ref{Eqn_asmpt})
\begin{eqnarray}
\lefteqn{\hat{\mathcal{V}}\Phi_{\infty}\left(  \mathbf{r}\right)    =}\nonumber\\
&  -\frac{\hbar^{2}}{2\mu}\sum_{lm}\beta_{lm}Y_{lm}\left(  \theta
,\varphi\right)  \frac{\left(  2l-1\right)  !!(l+1)}{k^{l+1}}\frac
{\delta\left(  r\right)  }{r^{l+2}}. \label{Eqn_VpsDef}%
\end{eqnarray}
Thus the original potential $V(\mathbf{r})$ is replaced by a sum over ``lumped''
multipole sources placed at $\mathbf{r}=0$.
To complete the construction of the pseudo-potential, we need to determine
coefficients $\beta_{lm}$ in terms of $\Phi\left(  \mathbf{r}\right)  $. First
we relate the integration constants $\alpha_{lm}$ and $\beta_{lm}$ by
requiring the complete solution $\Phi$ to be regular at $r=0$%
\begin{equation}
\beta_{lm}=\sum_{l^{\prime}m^{\prime}}\mathcal{K}_{lm}^{l^{\prime}m^{\prime}%
}\alpha_{l^{\prime}m^{\prime}} \label{Eqn_betaKalpha} \, .
\end{equation}
Here $\mathcal{K}_{lm}^{l^{\prime}m^{\prime}}$ are the elements of the
reactance (or K--) matrix  used to parameterize multi-channel
scattering~\cite{MotMas65}. It is worth noting that the entire
dependence of the pseudopotential on the original potential will be
encapsulated in matrix elements of the K-matrix.

Let us review some properties of the K-matrix. First it is related to more
familiar
scattering matrix $S$ via $S=(1+i\mathcal{K})(1-i\mathcal{K})^{-1}$ and
further to transmission or T-matrix through $S=1-iT$.  For low-energy
collisions $\mathcal{K}\approx-\frac{1}{2}T$. K-matrix is real and symmetric
\begin{equation}
\mathcal{K}_{lm}^{l^{\prime}m^{\prime}}\left(  k\right)  =\mathcal{K}_{l^{\prime
}m^{\prime}}^{lm}\left(  k\right) \, . \label{Eq_Ksym}
\end{equation}
For identical bosons (fermions) only
even (odd) partial waves need to be considered.  The K-matrix is
diagonal in $l$ and $m$ for spherically-symmetric potentials
\begin{equation}
\left[  \mathcal{K}_{\mathrm{sph}}\right]  _{lm}^{l^{\prime}m^{\prime}}%
=\delta_{ll^{\prime}}\delta_{mm^{\prime}}\tan\delta_{l} \, , \label{Eq_Ksph}%
\end{equation}
where $\delta_{l}$ is the phase shift for partial wave $l$.
Compared to spherically-symmetric case, {\em anisotropic} potentials additionally couple
different partial waves. For example, dipolar interactions, Eq.~(\ref{Eq_VDD}),
couple $s$ and $d$ waves so that $\left[  \mathcal{K}_{\mathrm{DD}}\right]
_{00}^{20}\neq0$ and quadrupolar interactions (\ref{Eq_VQQ}) couple $s$ and
$g$ waves. The scalar part  $V^\mathrm{sph}\left( r \right)$ of the
potential~(\ref{Eq_Vexpand})
assures non-vanishing$~\mathcal{K}_{00}^{00}$.
In practice, K-matrix for non-spherical interactions
may be found from a solution of coupled radial equations~\cite{Gel97}.
In particular, it may be shown that for potentials
parameterized by Eq.~(\ref{Eq_Vexpand}),
$\mathcal{K}_{lm}^{l^{\prime}m^{\prime}}\propto\delta_{mm^{\prime}}$,
i.e. the K-matrix is diagonal with respect to magnetic quantum numbers.
In Appendix, I derive some elements of K-matrix in the Born approximation
for dipolar and quadrupolar interaction.

At this point we related the integration constants $\alpha_{lm}$ and $\beta_{lm}$ via
elements of K-matrix, Eq.~(\ref{Eqn_betaKalpha}).
Further, in the low-energy limit $k r_{\mathrm c} \ll 1$, the integration
constants $\alpha_{lm}$ may be expressed in terms of $\Phi_{\infty}\left(
\mathbf{r}\right)  $ \cite{HuaYan57}
\begin{equation}
\alpha_{lm}=\frac{1}{2^{l}l!}\frac{1}{k^{l}}\left[  \left(  \frac{d}%
{dr}\right)  ^{2l+1}\left(  r^{l+1}\int Y_{lm}^{\ast}\left(  \Omega\right)
\Phi_{\infty}\left(  \mathbf{r}\right)  ~d\Omega\right)  \right]  _{r=0} \, .
\label{Eqn_alphaPhi}%
\end{equation}
Finally, combining equations (\ref{Eqn_VpsDef})--(\ref{Eqn_alphaPhi}) we arrive
at a generalization of pseudo-potential $\hat{\mathcal{V}}$ for anisotropic
interactions
\begin{eqnarray}
\hat{\mathcal{V}}\Phi\left(  \mathbf{r}\right)  =
&-\frac{\hbar^{2}}{M}%
\sum_{lm l^{\prime}m^{\prime}}\xi_{lm}^{l^{\prime}m^{\prime}}\left(
k\right)  ~\left(  \hat{v}_{lm}^{l^{\prime}m^{\prime}}\Phi\left(
\mathbf{r}\right)  \right) \, , \label{Eq_Vgps}
\end{eqnarray}
where
\begin{eqnarray}
\lefteqn{ \hat{v}_{lm}^{l^{\prime}m^{\prime}}\Phi\left(  \mathbf{r}\right)
=
\frac{\left(  2l\right)  !\left(  l+1\right)  }{2^{l+l^{\prime}}l^{\prime
}!l!} \, \, Y_{lm}\left(
\theta,\varphi\right)  \frac{\delta\left(  r\right)  }{r^{l+2}} \times} \nonumber \\
& &\left[  \left(  \frac{d}{dr}\right)  ^{2l^{\prime}+1}\left(
r^{l^{\prime}+1}\int Y_{l^{\prime}m^{\prime}}^{\ast}\left(  \Omega\right)
\Phi\left(  r,\Omega\right)  \, d\Omega\right)  \right]  _{r=0} \, , \\
\xi_{lm}^{l^{\prime}m^{\prime}}\left(  k\right)  &= &\frac{\mathcal{K}%
_{lm}^{l^{\prime}m^{\prime}}}{k^{l+l^{\prime}+1}} \, ,
\end{eqnarray}
and $M=2 \mu$ is a mass of the collision partner.

\section{Some properties of  pseudopotential}
\label{Sec_prop}
The derived anisotropic pseudopotential~(\ref{Eq_Vgps}) is one of the main results of this work.
The  spherically-symmetric pseudo-potential of \citet{HuaYan57} is subsumed in this equation.
Indeed, for spherically-symmetric interactions the K-matrix is diagonal in $l, l'$ and $m, m'$ and
is expressed in terms of conventional phase shifts, Eq.~(\ref{Eq_Ksph}). Upon substitution
of Eq.~(\ref{Eq_Ksph}) into pseudopotential~(\ref{Eq_Vgps}) we recover as a limiting case
the result of Ref.~\cite{HuaYan57}. As the pseudo-potential of \citet{HuaYan57}
the anisotropic pseudopotential is non-Hermitian and velocity dependent.

The pseudo-potential~(\ref{Eq_Vgps}) may be separated into spherically-symmetric
$\hat{\mathcal{V}}_\mathrm{sph}$ and
anisotropic  $\hat{\mathcal{V}}_\mathrm{anis}$ parts
\begin{equation}
\hat{\mathcal{V}} =
\hat{\mathcal{V}}_\mathrm{sph}  + \hat{\mathcal{V}}_\mathrm{anis} \, ,
\end{equation}
where
\begin{equation}
\hat{\mathcal{V}}_\mathrm{sph}  =
-\frac{\hbar^{2}}{M}%
\sum_{lm}\xi_{lm}^{lm}\left(
k\right) \,  \hat{v}_{lm}^{lm} \, \label{Eq_ps_sph}
\end{equation}
and
\begin{equation}
\hat{\mathcal{V}}_\mathrm{anis}  =
-\frac{\hbar^{2}}{M}%
\sum_{(lm) > (l^{\prime}m^{\prime})}\xi_{lm}^{l^{\prime}m^{\prime}}\left(
k\right) \,
\left( \hat{v}_{lm}^{l^{\prime}m^{\prime}} + \hat{v}^{lm}_{l^{\prime}m^{\prime}}\right) \, .
\label{Eq_ps_anis}
\end{equation}
In simplifying the anisotropic part we used symmetry property (\ref{Eq_Ksym}) of K-matrix.

Let us focus first on the spherically-symmetric part of the pseudopotential
and in the following section we will consider anisotropic part of the pseudopotential
for  dipolar and quadrupolar  interactions.
We expect that as in traditional BECs of dilute atomic gasses with spherically-symmetric
interactions, the effect of $\hat{\mathcal{V}}_\mathrm{sph}$ on
collective properties
will be determined mainly by s-wave contribution, i.e. $l=0,m=0$ term
in Eq.~(\ref{Eq_ps_sph})
\[
\hat{\mathcal{V}}_\mathrm{sph}  \approx
-\frac{\hbar^{2}}{M}  \frac{\mathcal{K}_{00}^{00}}{k}\hat{v}_{00}^{00} \, .
\]
It may be shown that for realistic potentials
the following low-energy limit is finite
\begin{equation}
a_{ss}=-\lim_{k\rightarrow0}\frac{\mathcal{K}_{00}^{00}\left(  k\right)  }%
{k} \, ; \label{Eq_ass}
\end{equation}
this quantity is a generalized scattering length. With this definition,
the truncated pseudopotential reduces to
\[
\hat{\mathcal{V}}_\mathrm{sph} \Phi\left(  \mathbf{r}\right)  \approx
4\pi~ \frac{\hbar^{2}}%
{M}a_{ss}~\delta\left(  \mathbf{r}\right)  \frac{\partial}{\partial
r}\left(  r~\Phi\left(  \mathbf{r}\right)  \right)  \text{,}%
\]
where we used $\delta\left(  \mathbf{r}\right)= \delta\left(
r\right)/ \left(  4\pi r^{2}\right)$ consistent with Ref.~\cite{HuaYan57}.
Finally, for sufficiently slowly-varying wavefunction,
$ |d \log \Phi / d \log r|  \ll 1$, we
recover conventional contact pseudo-potential
\begin{equation}
\hat{\mathcal{V}}_\mathrm{sph}  \approx 4\pi \, \frac{\hbar^{2}}%
{M}~a_{ss}~\delta\left(  \mathbf{r}\right) \, .
\label{Eq_contact}
\end{equation}
widely employed in studies of BECs.

Having discussed spherically-symmetric part of the pseudo-potential, in the following
section we consider anisotropic part of
pseudopotential~(\ref{Eq_ps_anis}) for dipolar and
quadrupolar  interactions.


\section{Dipolar and Quadrupolar interactions}
\label{Sec_DQ}
At this point we derived anisotropic pseudo-potential, Eq.(\ref{Eqn_VpsDef}).
We separated the pseudopotential into spherically-symmetric  and anisotropic contributions.
We found that the spherically-symmetric contribution reduces to familiar contact term (\ref{Eq_contact}),
widely employed in studies of Bose condensates; the only modification being an introduction
of generalized scattering length~(\ref{Eq_ass}).
In this section we focus on the anisotropic contribution and illustrate some of its properties
for dipolar and quadrupolar interactions of identical
bosons. These interactions were defined in the introductory Section as
potentials that at large separations $r$ are dominated by
\begin{equation}
V_\mathrm{anis}^\mathrm{DD}\left( r, \theta \right)  \rightarrow
\frac{C_3}{r^{3}}%
P_{2}\left(  \cos\theta\right)   \,,%
\end{equation}
for dipolar interactions
and
\begin{equation}
V_\mathrm{anis}^\mathrm{QQ}\left( r, \theta \right)
  \rightarrow  \frac{C_5}{r^{5}}%
P_{4}\left(  \cos\theta\right) \,  %
\end{equation}
for quadrupolar interactions.

Anisotropic interactions mix different partial waves $(lm)$ and $(l'm')$ via
off-diagonal elements  of K-matrix. From examining Eq.~(\ref{Eq_radial}) in Appendix, one
may determine that angular selection rules lead to
coupling of $s$ and $d$ waves for dipolar interactions and $s$ and $g$ waves for
quadrupolar interactions. In the following we assume that the dominant anisotropic effect on condensate
properties arises due to these particular couplings. Therefore,
\begin{eqnarray*}
\hat{\mathcal{V}}^\mathrm{DD}_\mathrm{anis} & \approx &-\frac{\hbar^{2}}%
{M}\, \frac
{\mathcal{K}_{00}^{20}}{k^{3}}\left(  \hat{v}_{00}^{20}+\hat{v}_{20}%
^{00}\right) \, , \\
\hat{\mathcal{V}}^\mathrm{QQ}_\mathrm{anis} & \approx &-\frac{\hbar^{2}}%
{M}\, \frac
{\mathcal{K}_{00}^{40}}{k^{5}}\left(  \hat{v}_{00}^{40}+\hat{v}_{40}%
^{00}\right) \, .
\end{eqnarray*}
For dipolar interactions it may be shown (Refs.~\cite{MarYou98,DebYou01} and Appendix of this paper)
that the following low-energy limit is finite
\begin{equation}
a^\mathrm{DD}_{sd}=-\lim_{k\rightarrow0}\frac{\mathcal{K}_{00}^{20}\left(  k\right)  }%
{k} \, . \label{Eq_asd}
\end{equation}
We will call this quantity off-diagonal scattering length.
Similarly, for quadrupolar interactions we may introduce off-diagonal scattering volume
\begin{equation}
a^\mathrm{QQ}_{sg}=-\lim_{k\rightarrow0}\frac{\mathcal{K}_{00}^{40}\left(  k\right)  }%
{k^3} \, . \label{Eq_asg}
\end{equation}
In the Appendix we employ Born approximation and find
\begin{eqnarray*}
a^\mathrm{DD}_{sd }& \approx & \frac{1}{12 \sqrt{5}} \frac{M}{\hbar^2}  C_3  \, ,\\
a^\mathrm{QQ}_{sg }& \approx & \frac{2}{7!} \frac{M}{\hbar^2}  C_5 \, .
\end{eqnarray*}
It is worth noting that the above results are valid only away from resonances.
In a general case one has to
find off-diagonal scattering length or volume by solving
corresponding scattering problem.

Finally the total truncated pseudopotential is given by
\begin{equation}
\hat{\mathcal{V}}  \approx 4\pi \, \frac{\hbar^{2}}%
{M}~a_{ss}~\delta\left(  \mathbf{r}\right) + \hat{\mathcal{V}}^\mathrm{DD,QQ}_\mathrm{anis}\, ,
\label{Eq_ps_sum}
\end{equation}
with
\begin{eqnarray}
\lefteqn{
\hat{\mathcal{V}}^\mathrm{DD}_\mathrm{anis}\Phi\left(  \mathbf{r}\right)  \approx
 \frac{\hbar^{2}}
{M}\frac{a_{sd}^\mathrm{DD}}{k^{2}} \sqrt{5} \times} \nonumber \\
&  \left\{ \frac{1}{8} \delta\left(  \mathbf{r}\right)
\left[  \left(  \frac{\partial}{\partial r}\right)
^{5}r^{3}\int P_{2}\left(  \cos\theta\right)  \Phi\left(  \mathbf{r}\right)
d\Omega\right]  +\right. \nonumber \\
&  \left. 9 \frac{\delta\left(  r\right)  }{r^{4}}P_{2}\left(  \cos
\theta\right)  \left[  \frac{\partial}{\partial r}\left(
r~\Phi\left(  \mathbf{r}\right)  \right)  \right]  _{r\rightarrow0}\right\}
\label{Eq_Vps_DD}
\end{eqnarray}
for dipolar interactions and
\begin{eqnarray}
\lefteqn{
\hat{\mathcal{V}}^\mathrm{QQ}_\mathrm{anis}\Phi\left(  \mathbf{r}\right)  \approx
 \frac{1}{2^7 }  \frac{\hbar^{2}}{M}\frac{a_{sg}^\mathrm{QQ}}{k^{2}} \times}
\nonumber  \\
& \left\{ \delta\left( \mathbf{ r} \right)
\left[  \left(  \frac{\partial}{\partial r}\right)
^{9}r^{5}\int P_{4}\left(  \cos\theta\right)  \Phi\left(  \mathbf{r}\right)
d\Omega\right]  +\right.  \nonumber \\
&  \left. 5 \, (8!) \frac{\delta\left(  r\right)}{r^6} P_{4}\left(  \cos
\theta\right)  \left[  \frac{\partial}{\partial r}\left(
r~\Phi\left(  \mathbf{r}\right)  \right)  \right]  _{r\rightarrow0}\right\}
\label{Eq_Vps_QQ}
\end{eqnarray}
for quadrupolar interactions. Quantities $\delta(r)/r^n$ may be recognized
as $n^\mathrm{th}$ derivatives of the Dirac delta-function.

The constructed pseudo-potential depends on the relative momentum $k$, i.e.
the potential is velocity dependent. In practice, the velocity-dependence
is most easily treated in momentum representation  and
in the next section we evaluate matrix element of the derived
pseudo-potential in the  basis of plane waves.

\section{Matrix element of anisotropic pseudo-potential in free-particle  basis}
\label{Sec_mel}
While considering effects of two-particle interactions on properties of a quantum
many-body system, one may require a matrix element of the derived pseudo-potential
in free-particle (plane-wave) basis.  We define this matrix element as
\begin{eqnarray}
\lefteqn{\bar{V}\left(  \mathbf{k}_{1},\mathbf{k}_{2},\mathbf{k}_{1}^{\prime
},\mathbf{k}_{2}^{\prime}\right)  \equiv } \label{Eqn_DefG} \\
& \frac{1}{\left(  2\pi\right)  ^{6}%
}\int d\mathbf{r\ }d\mathbf{r}^{\prime}e^{-i\mathbf{k}_{1}\mathbf{\cdot
r}}e^{-i\mathbf{k}_{2}\mathbf{\cdot r}^{\prime}}\hat{\mathcal{V}}\left(
\mathbf{r}-\mathbf{r}^{\prime}\right)  e^{i\mathbf{k}_{1}^{\prime}%
\cdot\mathbf{r}}e^{i\mathbf{k}_{2}^{\prime}\cdot\mathbf{r}^{\prime}%
} \, . \nonumber%
\end{eqnarray}
The pseudo-potential $\hat{\mathcal{V}}$, Eq.~(\ref{Eq_Vgps}), depends on the momentum of
relative motion of the interacting pair $\mathbf{k}=\frac{1}{2}\left(
\mathbf{\hat{p}}-\mathbf{\hat{p}}^{\prime}\right)  $, where $\mathbf{\hat{p}}$
and $\mathbf{\hat{p}}^{\prime}$ are  momenta conjugated to $\mathbf{r}$ and
$\mathbf{r}^{\prime}$ respectively.  To separate center of mass and relative
motions, we change the variables to  $\mathbf{R}\mathbf{=}\left(
\mathbf{r}+\mathbf{r}^{\prime}\right)  /2$ and $\mathbf{r}_{12}\mathbf{=r-r}%
^{\prime}$. With such a substitution
\begin{eqnarray*}
\lefteqn{\bar{V}\left(  \mathbf{k}_{1},\mathbf{k}_{2},\mathbf{k}_{1}^{\prime
},\mathbf{k}_{2}^{\prime}\right)   =
  \frac{1}{\left(  2\pi\right)  ^{3}} \, \delta_{\mathbf{k}_{1}^{\prime
}+\mathbf{k}_{2}^{\prime},\mathbf{k}_{1}+\mathbf{k}_{2}}\left(  -\frac
{\hbar^{2}}{M}\right) \times }\\
&&
 \sum_{ll^{\prime}mm^{\prime}}
 \xi_{lm}^{l^{\prime}m^{\prime}}\left(  k^{\prime}\right)  \int d\mathbf{r}_{12}%
\exp\left[  -i\mathbf{k}\cdot\mathbf{r}_{12}\right]  \hat{v}_{lm}^{l^{\prime
}m^{\prime}}\exp\left[  i\mathbf{k}^{\prime}\mathbf{\cdot~r}_{12}\right]  ~.
\end{eqnarray*}
Here we introduced two relative momenta
\begin{equation}
\mathbf{k}=\frac{1}{2}\left(  \mathbf{k}_{1}-\mathbf{k}_{2}\right)
~\mathrm{and~}\mathbf{k}^{\prime}\mathbf{=}\frac{1}{2}\left(  \mathbf{k}%
_{1}^{\prime}-\mathbf{k}_{2}^{\prime}\right)  ~.\label{Eqn_krel}%
\end{equation}
The delta function $\delta_{\mathbf{k}_{1}^{\prime}+\mathbf{k}_{2}^{\prime
},~\mathbf{k}_{1}+\mathbf{k}_{2}}$ ensures conservation of the total linear
momentum. Further we use partial-wave expansion
\[
\exp\left[  i\mathbf{k}^{\prime}\mathbf{\cdot~r}_{12}\right]  =4\pi\sum
_{l_{1}m_{1}}i^{l_{1}}j_{l_{1}}\left(  k^{\prime}r_{12}\right)  Y_{l_{1}m_{1}%
}^{\ast}\left(  \hat{k}^{\prime}\right)  Y_{l_{1}m_{1}}\left(  \hat{r}%
_{12}\right)
\]
and arrive at%
\begin{eqnarray*}
\lefteqn{
\int d\mathbf{r}_{12}\exp\left[  -i\mathbf{k}\cdot\mathbf{r}_{12}\right]
\hat{v}_{lm}^{l^{\prime}m^{\prime}}\left(  \mathbf{r}_{12}\right)  \exp\left[
i\mathbf{k}^{\prime}\cdot \mathbf{ r}_{12}\right]  = } \\
& &
(4\pi)^{2}~i^{l^{\prime}-l}\frac{l+1}{2l+1}\left(  k^{\prime}\right)^{l^{\prime}}k^{l}~Y_{l^{\prime
}m^{\prime}}^{\ast}\left(  \hat{k}^{\prime}\right)
Y_{lm}\left(  \hat{k}\right) \, .%
\end{eqnarray*}
Finally, the matrix element of the anisotropic pseudo-potential
may be expressed in terms of
relative momenta as
\begin{eqnarray*}
\bar{V}\left(  \mathbf{k}_{1},\mathbf{k}_{2},\mathbf{k}_{1}^{\prime
},\mathbf{k}_{2}^{\prime}\right)  &=&
\delta_{\mathbf{k}_{1}^{\prime}+\mathbf{k}_{2}^{\prime},\mathbf{k}_{1}+\mathbf{k}_{2}} \times \nonumber \\
& & \bar{v}\left(
\frac{1}{2}\left(  \mathbf{k}_{1}-\mathbf{k}_{2}\right), \frac{1}{2}\left(
\mathbf{k}_{1}^{\prime}-\mathbf{k}_{2}^{\prime}\right)  \right)  \, ,
\end{eqnarray*}
with
\begin{eqnarray}
\bar{v}\left(  \mathbf{k},\mathbf{k}^{\prime}\right)  &=& -\frac{\hbar^{2}}
{M}\frac{1}{\pi}\sum_{ll^{\prime}mm^{\prime}}i^{l^{\prime}-l}
\frac{\mathcal{K}_{lm}^{l^{\prime}m^{\prime}}\left(  k^{\prime}\right)
}{k^{\prime}} \times\nonumber \\
& & \left(  \frac{k}{k^{\prime}}\right)^{l}
\frac{2l+2}{2l+1}
Y_{l^{\prime}m^{\prime}}^{\ast}\left(  \hat{k}^{\prime}\right)
Y_{lm}\left(  \hat{k}\right)  \, . \label{Eqn_vbar}
\end{eqnarray}

Let us once again specialize the discussion to dipolar and quadrupolar
interactions. As in Section~\ref{Sec_DQ} we assume that the dominant anisotropic effect
arises due to couplings of $s$ and $d$ partial waves
for dipolar interactions and due to mixing of $s$ and $g$ waves for quadrupolar
interactions. The corresponding truncated  matrix element
(\ref{Eqn_vbar}) may be represented as
\begin{equation}
\bar{v}\left(  \mathbf{k},\mathbf{k}^{\prime}\right)  \approx \frac{1}{2\pi^{2}}%
\frac{\hbar^{2}}{M}\left\{  a_{ss}+\mathcal{F}\left(  \mathbf{k}%
,\mathbf{k}^{\prime}\right)  \right\}  \label{Eqn_GDD}
\end{equation}
with $\mathcal{F}$ replaced by
\begin{eqnarray*}
\mathcal{F}^\mathrm{DD}\left(  \mathbf{k},\mathbf{k}^{\prime}\right)  =
- a_{sd}^\mathrm{DD}\left\{ \sqrt{5}%
P_{2}\left(  \cos\theta_{k^{\prime}}\right)+\frac{3}{\sqrt{5}}\left(  \frac
{k}{k^{\prime}}\right)  ^{2}P_{2}\left(  \cos\theta_{k}\right)
\, \right\}
\end{eqnarray*}
for dipolar interactions and by
\begin{eqnarray*}
\mathcal{F}^\mathrm{QQ}\left(  \mathbf{k},\mathbf{k}^{\prime}\right)  =
a_{sg}^\mathrm{QQ}\left\{
3 (k')^2 P_4( \cos \theta_{k'} ) +
 \frac{10}{3} k^2 \left(  \frac{k}{k^{\prime}}\right)^{2} P_{4}\left(  \cos\theta_{k}\right)
\, \right\}
\end{eqnarray*}
for quadrupolar interactions.
In these expressions $a_{ss}$ is a generalized scattering length~(\ref{Eq_ass}) and
$a_{sd}^\mathrm{DD}$ and $a_{sg}^\mathrm{QQ}$ are off-diagonal scattering length and
volume defined by Eq.~(\ref{Eq_asd}) and Eq.~(\ref{Eq_asg}) respectively.

\section{Conclusion}
I rigorously derived anisotropic pseudopotential
arising in the context of adiabatic collisions of two particles polarized by
external field. Such low-energy pseudopotential may be useful in describing
collective properties of dilute quantum gases, such as molecules polarized
by an external electric field or metastable $^3P_2$ atoms polarized by
magnetic field.
The pseudopotential is given by Eq.(\ref{Eq_Vgps}).
It is naturally expressed in terms of reactance (K--) matrix.
The potential is non-Hermitian and velocity-dependent.
It worth noting that in the derivation I did not require the
validity of the Born approximation as in Ref.~\cite{YiYou00}.
Rather I followed method of \citet{HuaYan57} and
at large separations demanded the equality of solutions of two-body Schr\"{o}dinger equation
with a full original potential and with a pseudo-potential. Thus,
compared to Eq.~(\ref{Eqn_YY}) by \citet{YiYou00}, the derived two-body
pseudopotential is expected to be also valid in a vicinity of scattering resonances.

I argued that in most applications the pseudopotential may be represented
as a sum of traditional spherically-symmetric contact term and
anisotropic part, Eq.~(\ref{Eq_ps_sum}). The former contribution may
be parameterized by a generalized scattering length~(\ref{Eq_ass}).
We specialized discussion of the anisotropic part of pseudopotential
to dipolar and quadrupolar interactions and
found that it
can be characterized by
off-diagonal scattering length $a_{sd}^\mathrm{DD}$, Eq.~(\ref{Eq_asd}), for dipolar
interactions and off-diagonal scattering volume $a_{sg}^\mathrm{QQ}$, Eq.~(\ref{Eq_asg}), for quadrupolar
interactions. Although in a particular application these parameters should be determined
from a solution of multi-channel scattering problem, I have derived $a_{sd}^\mathrm{DD}$
and $a_{sg}^\mathrm{QQ}$ in the Born approximations.
Keeping in mind many-body applications, I have also derived two-body matrix-element in the
plane-wave basis, Eq.~(\ref{Eqn_vbar}). Thus in this work I have rigorously
derived anisotropic pseudo-potential for polarized dilute quantum gases and investigated
its properties.

\begin{acknowledgments}
I would like to thank Eite Tiesinga for stimulating discussions
and Li You and Su Yi for comments on manuscript.
This work was supported in part by the National Science Foundation.
\end{acknowledgments}

\appendix
\section{Off-diagonal scattering length $a^\mathrm{DD}_{sd}$ and
scattering volume $a^\mathrm{QQ}_{sg}$ in the Born approximation
}
Here  we obtain expressions for reactance (K--) matrix in the Born approximation.
Using the derived K-matrix we estimate  off-diagonal scattering length $a^\mathrm{DD}_{sd}$ and
scattering volume $a^\mathrm{QQ}_{sg}$ for dipolar (DD) and quadrupolar (QQ) interactions.

The full solution of the  Schr\"{o}dinger equation~(\ref{Eq_SE}) may be represented as
\begin{equation}
\Phi\left(  \mathbf{r} \right)=
\sum_{lm}Y_{lm}\left(  \theta,\phi\right)
 \frac{u_{lm}\left(  r\right) }{k r} \, .
\end{equation}
It can be shown that the radial functions $u_{lm}\left(  r\right)$
satisfy the following system of coupled differential equations
\begin{eqnarray}
\lefteqn{\left\{  \frac{d^{2}}{dr^{2}}-\frac{l\left(  l+1\right)  }{r^{2}}%
+k^{2}\right\}  u_{lm}\left(  r\right)  =} \nonumber \\
& &2\mu\sum_{l^{\prime}m^{\prime}%
}~\langle lm|V|l^{\prime}m^{\prime}\rangle u_{l^{\prime}m^{\prime}}\left(
r\right) \label{Eq_radial} ,
\end{eqnarray}
with%
\begin{equation}
\langle lm|V|l^{\prime}m^{\prime}\rangle\left(  r\right)  =\int d\Omega
~Y_{lm}^{\ast}\left(  \Omega\right)  V\left(  \mathbf{r}\right)  Y_{l^{\prime
}m^{\prime}}\left(  \Omega\right) \, .
\end{equation}
It is convenient to introduce regular and irregular solutions of
homogeneous radial equations
\begin{eqnarray*}
F_{l}\left(  kr\right)    & = & kr \, j_{l}\left(  kr\right) \, , \\
G_{l}\left(  kr\right)    & =& -kr \, n_{l}\left(  kr\right)
\end{eqnarray*}
and corresponding  standing-wave Green's function
\[
g_{l}\left(  r,r^{\prime}\right)  =  -\frac{1}{k}
\left\{
\begin{array}{cc}
  G_{l}\left(  kr^{\prime}\right)  F_{l}\left( kr\right) \, , & r<r^{\prime} \\
  G_{l}\left(  k r\right)  F_{l}\left(  k r^{\prime}\right) \,,  & r>r^{\prime} \\
\end{array}
\right. \, .
\]
Using these definitions, solutions to the system of
inhomogenous equations~(\ref{Eq_radial}) regular at $r=0$
may be represented as
\begin{equation}
u_{lm}\left(  r\right)  =\alpha_{lm}F_{l}\left(  kr\right)  +\int_{0}^{\infty
}dr^{\prime}g_{l}\left(  r,r^{\prime}\right)  \left[  2\mu\sum_{l^{\prime
}m^{\prime}}~\langle lm|V|l^{\prime}m^{\prime}\rangle u_{l^{\prime}m^{\prime}%
}\left(  r^{\prime}\right)  \right]  \,, \label{Eq_integral}
\end{equation}
where constants $\alpha_{lm}$ are chosen to satisfy some boundary conditions.
In the spirit of Born approximation  we may find solution
of integral equations (\ref{Eq_integral})
iteratively
starting from
\[
u_{lm}\left(  r\right)  \approx \alpha_{lm}F_{l}\left( kr \right) \, .
\]
In the lowest order in $V$ at large $r$ one obtains
\[
u_{lm}\left(  r\right)  \rightarrow\alpha_{lm}F_{l}\left(  kr\right)
+G_{l}\left(  kr\right)  \alpha_{lm}\sum_{l^{\prime}m^{\prime}}\left[
-\frac{2\mu}{k}\int_{0}^{\infty}dr^{\prime}F_{l}\left(  kr^{\prime}\right)
~\langle lm|V|l^{\prime}m^{\prime}\rangle F_{l}\left(  kr^{\prime}\right)
\right] \, .
\]
By comparing with Eq.~(\ref{Eqn_asmpt}) and Eq.~(\ref{Eqn_betaKalpha}),
we arrive at expression for elements of
K-matrix in the Born approximation
\begin{equation}
\mathcal{K}_{lm}^{l^{\prime}m^{\prime}}\approx-\frac{2\mu k}{\hbar^{2}}\int_{0}^{\infty
}r^{2}~j_{l}\left(  kr\right)  ~j_{l^{\prime}}\left(  kr\right)  ~\langle
lm|V|l^{\prime}m^{\prime}\rangle dr \, .
\label{Eq_KmatBorn}%
\end{equation}

The Born approximation generally does not hold for low-energy atomic collisions, since
realistic interactions are singular at small $r$.
However, for dipolar interactions You and co-workers~\cite{MarYou98,DebYou01}
found numerically  that away from resonances Born approximation works
well for off-diagonal matrix elements. Keeping this observation in mind, below
we derive  off-diagonal scattering length and volume introduced in the main body of the paper.
These parameters were defined as low-energy limits
\begin{eqnarray*}
a^\mathrm{DD}_{sd }&=& -\lim_{k\rightarrow 0}\frac{[\mathcal{K}^\mathrm{DD}]_{00}^{20}\left(  k\right)  }%
{k} \,  , \\
a^\mathrm{QQ}_{sg }&=& -\lim_{k\rightarrow 0}\frac{[\mathcal{K}^\mathrm{QQ}]_{00}^{40}\left(  k\right)  }%
{k^3} \,
\end{eqnarray*}
for dipolar (DD) and quadrupolar (QQ) interactions.
In Section~\ref{Sec_intro}  the DD and QQ interactions were parameterized as
\begin{eqnarray*}
V_\mathrm{DD} &=&
\frac{C_3}{r^{3}}%
P_{2}\left(  \cos\theta\right) \, , \\
V_\mathrm{QQ} &=& \frac{C_5}{r^{5}}%
P_{4}\left(  \cos\theta\right) \, .
\end{eqnarray*}
Using these definitions and Eq.~(\ref{Eq_KmatBorn}), we arrive at
\begin{eqnarray}
a^\mathrm{DD}_{sd }&=& \frac{1}{6 \sqrt{5}} \frac{ \mu}{\hbar^2}  C_3  \, ,\\
a^\mathrm{QQ}_{sg }&=& \frac{1}{1260} \frac{ \mu}{\hbar^2}  C_5 \, .
\end{eqnarray}
It is worth emphasizing that these results were derived in the Born approximation.
In general case, to obtain parameters entering anisotropic pseudo-potential (\ref{Eq_Vgps}) one has
to  numerically solve the system of equations (\ref{Eq_radial}),
especially in the vicinity of resonances.

\bibliography{nonsphBEC,mypub,general,BEC}

\begin{thebibliography}{23}
\expandafter\ifx\csname natexlab\endcsname\relax\def\natexlab#1{#1}\fi
\expandafter\ifx\csname bibnamefont\endcsname\relax
  \def\bibnamefont#1{#1}\fi
\expandafter\ifx\csname bibfnamefont\endcsname\relax
  \def\bibfnamefont#1{#1}\fi
\expandafter\ifx\csname citenamefont\endcsname\relax
  \def\citenamefont#1{#1}\fi
\expandafter\ifx\csname url\endcsname\relax
  \def\url#1{\texttt{#1}}\fi
\expandafter\ifx\csname urlprefix\endcsname\relax\def\urlprefix{URL }\fi
\providecommand{\bibinfo}[2]{#2}
\providecommand{\eprint}[2][]{\url{#2}}

\bibitem[{\citenamefont{Demkov and Ostrovsky}(1988)}]{DemOst88}
\bibinfo{author}{\bibfnamefont{Y.~N.} \bibnamefont{Demkov}} \bibnamefont{and}
  \bibinfo{author}{\bibfnamefont{V.~N.} \bibnamefont{Ostrovsky}},
  \emph{\bibinfo{title}{Zero-range Potentials Method in Atomic Physics}}
  (\bibinfo{publisher}{Plenum}, \bibinfo{address}{New York},
  \bibinfo{year}{1988}).

\bibitem[{\citenamefont{Dalfovo et~al.}(1999)\citenamefont{Dalfovo, Giorgini,
  Pitaevskii, and Stringari}}]{DalGioPit99}
\bibinfo{author}{\bibfnamefont{F.}~\bibnamefont{Dalfovo}},
  \bibinfo{author}{\bibfnamefont{S.}~\bibnamefont{Giorgini}},
  \bibinfo{author}{\bibfnamefont{L.~P.} \bibnamefont{Pitaevskii}},
  \bibnamefont{and}
  \bibinfo{author}{\bibfnamefont{S.}~\bibnamefont{Stringari}},
  \bibinfo{journal}{Rev.\ Mod.\ Phys.} \textbf{\bibinfo{volume}{71}},
  \bibinfo{pages}{463} (\bibinfo{year}{1999}).

\bibitem[{\citenamefont{Leggett}(2001)}]{Leg01}
\bibinfo{author}{\bibfnamefont{A.~J.} \bibnamefont{Leggett}},
  \bibinfo{journal}{Rev.\ Mod.\ Phys.} \textbf{\bibinfo{volume}{73}},
  \bibinfo{pages}{307} (\bibinfo{year}{2001}).

\bibitem[{\citenamefont{Huang and Yang}(1957)}]{HuaYan57}
\bibinfo{author}{\bibfnamefont{K.}~\bibnamefont{Huang}} \bibnamefont{and}
  \bibinfo{author}{\bibfnamefont{C.~N.} \bibnamefont{Yang}},
  \bibinfo{journal}{Phys.\ Rev.\ A} \textbf{\bibinfo{volume}{105}},
  \bibinfo{pages}{767} (\bibinfo{year}{1957}).

\bibitem[{\citenamefont{Yi and You}(2000)}]{YiYou00}
\bibinfo{author}{\bibfnamefont{S.}~\bibnamefont{Yi}} \bibnamefont{and}
  \bibinfo{author}{\bibfnamefont{L.}~\bibnamefont{You}},
  \bibinfo{journal}{Phys. Rev. A} \textbf{\bibinfo{volume}{63}},
  \bibinfo{pages}{053607/1} (\bibinfo{year}{2000}).

\bibitem[{\citenamefont{Yi and You}(2002)}]{YiYou02}
\bibinfo{author}{\bibfnamefont{S.}~\bibnamefont{Yi}} \bibnamefont{and}
  \bibinfo{author}{\bibfnamefont{L.}~\bibnamefont{You}},
  \bibinfo{journal}{Phys. Rev. A} \textbf{\bibinfo{volume}{66}},
  \bibinfo{pages}{013607/1} (\bibinfo{year}{2002}).

\bibitem[{\citenamefont{Giovanazzi
  et~al.}(2002{\natexlab{a}})\citenamefont{Giovanazzi, G\"{o}rlitz, and
  Pfau}}]{GioGorPfa02}
\bibinfo{author}{\bibfnamefont{S.}~\bibnamefont{Giovanazzi}},
  \bibinfo{author}{\bibfnamefont{A.}~\bibnamefont{G\"{o}rlitz}},
  \bibnamefont{and} \bibinfo{author}{\bibfnamefont{T.}~\bibnamefont{Pfau}},
  \bibinfo{journal}{Phys. Rev. Lett.} \textbf{\bibinfo{volume}{89}},
  \bibinfo{pages}{130401} (\bibinfo{year}{2002}{\natexlab{a}}).

\bibitem[{\citenamefont{Santos et~al.}(2000)\citenamefont{Santos, Shlyapnikov,
  Zoller, and Lewenstein}}]{SanShlZol00}
\bibinfo{author}{\bibfnamefont{L.}~\bibnamefont{Santos}},
  \bibinfo{author}{\bibfnamefont{G.~V.} \bibnamefont{Shlyapnikov}},
  \bibinfo{author}{\bibfnamefont{P.}~\bibnamefont{Zoller}}, \bibnamefont{and}
  \bibinfo{author}{\bibfnamefont{M.}~\bibnamefont{Lewenstein}},
  \bibinfo{journal}{Phys. Rev. Lett.} \textbf{\bibinfo{volume}{85}},
  \bibinfo{pages}{1791} (\bibinfo{year}{2000}).

\bibitem[{\citenamefont{G\'{o}ral et~al.}(2000)\citenamefont{G\'{o}ral,
  Rz\c{a}\.{z}ewski, and Pfau}}]{GorRzaPfa00}
\bibinfo{author}{\bibfnamefont{K.}~\bibnamefont{G\'{o}ral}},
  \bibinfo{author}{\bibfnamefont{K.}~\bibnamefont{Rz\c{a}\.{z}ewski}},
  \bibnamefont{and} \bibinfo{author}{\bibfnamefont{T.}~\bibnamefont{Pfau}},
  \bibinfo{journal}{Phys. Rev. A} \textbf{\bibinfo{volume}{61}},
  \bibinfo{pages}{051601/1} (\bibinfo{year}{2000}).

\bibitem[{\citenamefont{G\'{o}ral and Santos}(2002)}]{GorSan02}
\bibinfo{author}{\bibfnamefont{K.}~\bibnamefont{G\'{o}ral}} \bibnamefont{and}
  \bibinfo{author}{\bibfnamefont{L.}~\bibnamefont{Santos}},
  \bibinfo{journal}{Phys. Rev. A} \textbf{\bibinfo{volume}{66}},
  \bibinfo{pages}{023613} (\bibinfo{year}{2002}).

\bibitem[{\citenamefont{G\'{o}ral et~al.}(2002)\citenamefont{G\'{o}ral, Santos,
  and Lewenstein}}]{GorSanLew02}
\bibinfo{author}{\bibfnamefont{K.}~\bibnamefont{G\'{o}ral}},
  \bibinfo{author}{\bibfnamefont{L.}~\bibnamefont{Santos}}, \bibnamefont{and}
  \bibinfo{author}{\bibfnamefont{M.}~\bibnamefont{Lewenstein}},
  \bibinfo{journal}{Phys. Rev. Lett.} \textbf{\bibinfo{volume}{88}},
  \bibinfo{pages}{170406} (\bibinfo{year}{2002}).

\bibitem[{\citenamefont{Giovanazzi
  et~al.}(2002{\natexlab{b}})\citenamefont{Giovanazzi, O'Dell, and
  Kurizki}}]{GioODeKur02}
\bibinfo{author}{\bibfnamefont{S.}~\bibnamefont{Giovanazzi}},
  \bibinfo{author}{\bibfnamefont{D.}~\bibnamefont{O'Dell}}, \bibnamefont{and}
  \bibinfo{author}{\bibfnamefont{G.}~\bibnamefont{Kurizki}},
  \bibinfo{journal}{Phys. Rev. Lett.} \textbf{\bibinfo{volume}{88}},
  \bibinfo{pages}{130402} (\bibinfo{year}{2002}{\natexlab{b}}).

\bibitem[{\citenamefont{G\'{o}ral et~al.}(2001)\citenamefont{G\'{o}ral,
  Englert, and Rz\c{a}\.{z}ewski}}]{KrzEngRze01}
\bibinfo{author}{\bibfnamefont{K.}~\bibnamefont{G\'{o}ral}},
  \bibinfo{author}{\bibfnamefont{B.-G.} \bibnamefont{Englert}},
  \bibnamefont{and}
  \bibinfo{author}{\bibfnamefont{K.}~\bibnamefont{Rz\c{a}\.{z}ewski}},
  \bibinfo{journal}{Phys. Rev. A} \textbf{\bibinfo{volume}{63}},
  \bibinfo{pages}{033606} (\bibinfo{year}{2001}).

\bibitem[{\citenamefont{Baranov et~al.}(2002)\citenamefont{Baranov, Mar'enko,
  Rychkov, and Shlyapnikov}}]{BarMarRyc02}
\bibinfo{author}{\bibfnamefont{M.~A.} \bibnamefont{Baranov}},
  \bibinfo{author}{\bibfnamefont{M.~S.} \bibnamefont{Mar'enko}},
  \bibinfo{author}{\bibfnamefont{V.~S.} \bibnamefont{Rychkov}},
  \bibnamefont{and} \bibinfo{author}{\bibfnamefont{G.~V.}
  \bibnamefont{Shlyapnikov}}, \bibinfo{journal}{Phys. Rev. A}
  \textbf{\bibinfo{volume}{66}}, \bibinfo{pages}{013606}
  (\bibinfo{year}{2002}).

\bibitem[{\citenamefont{Kim et~al.}(1997)\citenamefont{Kim, Friedrich, Katz,
  Patterson, Weinstein, DeCarvalho, and Doyle}}]{KimFriKat97}
\bibinfo{author}{\bibfnamefont{J.}~\bibnamefont{Kim}},
  \bibinfo{author}{\bibfnamefont{B.}~\bibnamefont{Friedrich}},
  \bibinfo{author}{\bibfnamefont{D.~P.} \bibnamefont{Katz}},
  \bibinfo{author}{\bibfnamefont{D.}~\bibnamefont{Patterson}},
  \bibinfo{author}{\bibfnamefont{J.~D.} \bibnamefont{Weinstein}},
  \bibinfo{author}{\bibfnamefont{R.}~\bibnamefont{DeCarvalho}},
  \bibnamefont{and} \bibinfo{author}{\bibfnamefont{J.~M.} \bibnamefont{Doyle}},
  \bibinfo{journal}{Phys. Rev. Lett.} \textbf{\bibinfo{volume}{78}},
  \bibinfo{pages}{3665} (\bibinfo{year}{1997}).

\bibitem[{\citenamefont{Weinstein et~al.}(1998)\citenamefont{Weinstein,
  deCarvalho, Kim, Patterson, Friedrich, and Doyle}}]{WeideCKim98}
\bibinfo{author}{\bibfnamefont{J.~D.} \bibnamefont{Weinstein}},
  \bibinfo{author}{\bibfnamefont{R.}~\bibnamefont{deCarvalho}},
  \bibinfo{author}{\bibfnamefont{J.}~\bibnamefont{Kim}},
  \bibinfo{author}{\bibfnamefont{D.}~\bibnamefont{Patterson}},
  \bibinfo{author}{\bibfnamefont{B.}~\bibnamefont{Friedrich}},
  \bibnamefont{and} \bibinfo{author}{\bibfnamefont{J.~M.} \bibnamefont{Doyle}},
  \bibinfo{journal}{Phys. Rev. A} \textbf{\bibinfo{volume}{57}},
  \bibinfo{pages}{R3173} (\bibinfo{year}{1998}).

\bibitem[{\citenamefont{Bell et~al.}(1999)\citenamefont{Bell, Stuhler, Locher,
  Hensler, Mlynek, and Pfau}}]{BelStuLoc99}
\bibinfo{author}{\bibfnamefont{A.~S.} \bibnamefont{Bell}},
  \bibinfo{author}{\bibfnamefont{J.}~\bibnamefont{Stuhler}},
  \bibinfo{author}{\bibfnamefont{S.}~\bibnamefont{Locher}},
  \bibinfo{author}{\bibfnamefont{S.}~\bibnamefont{Hensler}},
  \bibinfo{author}{\bibfnamefont{J.}~\bibnamefont{Mlynek}}, \bibnamefont{and}
  \bibinfo{author}{\bibfnamefont{T.}~\bibnamefont{Pfau}},
  \bibinfo{journal}{Europhys. Lett.} \textbf{\bibinfo{volume}{45}},
  \bibinfo{pages}{156} (\bibinfo{year}{1999}).

\bibitem[{\citenamefont{Derevianko et~al.}()\citenamefont{Derevianko, Porsev,
  Kotochigova, Tiesinga, and Julienne}}]{DerPorKot03}
\bibinfo{author}{\bibfnamefont{A.}~\bibnamefont{Derevianko}},
  \bibinfo{author}{\bibfnamefont{S.~G.} \bibnamefont{Porsev}},
  \bibinfo{author}{\bibfnamefont{S.}~\bibnamefont{Kotochigova}},
  \bibinfo{author}{\bibfnamefont{E.}~\bibnamefont{Tiesinga}}, \bibnamefont{and}
  \bibinfo{author}{\bibfnamefont{P.~S.} \bibnamefont{Julienne}},
  \bibinfo{note}{e-print:physics/0210076}.

\bibitem[{\citenamefont{Fetter and Walecka}(1971)}]{FetWal71}
\bibinfo{author}{\bibfnamefont{A.~L.} \bibnamefont{Fetter}} \bibnamefont{and}
  \bibinfo{author}{\bibfnamefont{J.~D.} \bibnamefont{Walecka}},
  \emph{\bibinfo{title}{Quantum Theory of Many-particle Systems}}
  (\bibinfo{publisher}{McGraw-Hill}, \bibinfo{year}{1971}).

\bibitem[{\citenamefont{Mott and Massey}(1965)}]{MotMas65}
\bibinfo{author}{\bibfnamefont{N.~F.} \bibnamefont{Mott}} \bibnamefont{and}
  \bibinfo{author}{\bibfnamefont{H.~S.~W.} \bibnamefont{Massey}},
  \emph{\bibinfo{title}{The theory of atomic collisions}}
  (\bibinfo{publisher}{Oxford University Press}, \bibinfo{address}{London},
  \bibinfo{year}{1965}), \bibinfo{edition}{3rd} ed.

\bibitem[{\citenamefont{Geltman}(1997)}]{Gel97}
\bibinfo{author}{\bibfnamefont{S.}~\bibnamefont{Geltman}},
  \emph{\bibinfo{title}{Topics in atomic collision theory}}
  (\bibinfo{publisher}{Krieger Pub. Co.}, \bibinfo{address}{Florida},
  \bibinfo{year}{1997}).

\bibitem[{\citenamefont{Marinescu and You}(1998)}]{MarYou98}
\bibinfo{author}{\bibfnamefont{M.}~\bibnamefont{Marinescu}} \bibnamefont{and}
  \bibinfo{author}{\bibfnamefont{L.}~\bibnamefont{You}},
  \bibinfo{journal}{Phys. Rev. Lett.} \textbf{\bibinfo{volume}{81}},
  \bibinfo{pages}{4596} (\bibinfo{year}{1998}).

\bibitem[{\citenamefont{Deb and You}(2001)}]{DebYou01}
\bibinfo{author}{\bibfnamefont{B.}~\bibnamefont{Deb}} \bibnamefont{and}
  \bibinfo{author}{\bibfnamefont{L.}~\bibnamefont{You}},
  \bibinfo{journal}{Phys. Rev. A} \textbf{\bibinfo{volume}{64}},
  \bibinfo{pages}{022717/1} (\bibinfo{year}{2001}).

\end{thebibliography}

\end{document}